\documentstyle[12pt]{article}
\newcommand{\beq}{\begin{equation}}
\newcommand{\eeq}{\end{equation}}
\begin{document}

\begin{titlepage}
\begin{center}
{\hbox to\hsize{hep-th/9906064  \hfill   MIT-CTP-2874}} 

{\hbox to\hsize{ 
\hfill PUPT-1867}}

{\hbox to\hsize{ 
\hfill BUHEP-99-13}}

\bigskip
\vspace{3\baselineskip}

{\Large \bf  
 
An Alternative to Compactification\\}

\bigskip

\bigskip

{\bf  Lisa Randall}\\
\smallskip

{ \small \it  
Joseph Henry Laboratories,
Princeton University,
Princeton, NJ 08543, USA\\

and

 Center for Theoretical Physics,

Massachusetts Institute of Technology, Cambridge, MA 02139, USA }

\medskip
{\bf Raman Sundrum}\\
\smallskip
{\small \it Department of Physics, 

Boston University, Boston, MA 02215, USA}

\bigskip

{\tt  randall@baxter.mit.edu} \\
{\tt sundrum@budoe.bu.edu}

\bigskip

\vspace*{.5cm}

{\bf Abstract}\\
\end{center}
\noindent
Conventional wisdom states that  Newton's
force law implies only four non-compact dimensions.
We demonstrate that this is not necessarily true
in the presence of a non-factorizable background
geometry. The specific example we study
is a single 3-brane embedded in five dimensions.
We show that even without a gap in the Kaluza-Klein
spectrum, four-dimensional 
Newtonian and general relativistic gravity is reproduced 
to more than adequate precision.

\bigskip

\bigskip

\end{titlepage}

\section{Introduction}
There exists ``lore'' that
convinces us that we live in four non-compact
dimensions. Certainly Standard Model matter
cannot propagate a large distance in extra
dimensions without conflict with 
observations. As has recently been emphasized,  this
can be avoided if the Standard 
Model is confined to  
a $(3+1)$-dimensional subspace, or ``3-brane'',  in the higher dimensions 
[1 -- 8]. However, this solution will not work for  gravity,
which necessarily propagates in 
 all dimensions as it is the dynamics of spacetime itself. The experimental 
success of of Newton's $1/r^2$ law and 
general relativity would therefore seem to  imply  precisely four 
non-compact dimensions. 
Additional dimensions would be acceptable, so
long as they are  compact and sufficiently
small to be consistent with 
current gravitational tests. 
One further  piece of lore
is that if there are  $n$ extra compact dimensions,
the Planck scale is related to the higher
dimensional scale of gravity, $M$, through
the relation $M_{Pl}^2=M^{2+n} V_n$,
where $V_n$ is the extra-dimensional volume.

The point of this letter is to argue
that none of the statements about gravity  in the previous paragraph
is necessarily true. The
previous properties  rely on a factorizable
geometry, namely the metric
of the four familiar dimensions  is independent
of  coordinate  in the extra dimensions.
The story can change significantly 
when this assumption is dropped. Perhaps
the most dramatic consequence is
that we can live in $4+n$ {\it non-compact}
dimensions, in perfect compatibility with experimental 
gravity. We will give an example with
one additional dimension ($n=1$).
We will show that $M_{Pl}$ 
%and the masses of the Kaluza-Klein excitations of
% the graviton 
is  determined
by the higher-dimensional curvature rather than the size of the extra
dimension. This curvature is not in conflict  with four-dimensional Poincare
invariance. Earlier work on non-compact 
extra dimensions studied trapping of matter fields to be effectively 
four-dimensional \cite{shap}  or studied 
finite-volume but topologically non-compact extra dimensions \cite{zwi}.
 Here we carefully study the trapping of gravity itself to be 
effectively four-dimensional, where the extra-dimensional volume is  
infinite.

The reason the above statements can be true is that  a curved
background can support a ``bound
state'' of the higher-dimensional graviton, 
which is localized in the extra dimensions.  
So although space is indeed infinite in extent, the graviton
is confined to a small region within this space.
The existence of a bound state can be  understood as follows.
Small gravitational fluctuations satisfy a wave
equation of the form 
\begin{equation}
\left(\partial_\mu \partial^\mu -\partial_j \partial^j 
+ V(z_j)\right) \hat{h}(x^{\mu},z_j)=0,
\end{equation}
with a non-trivial ``potential'', $V$, arising from the curvature.
The $\mu$ indices run from 0 to 3 whereas the $j$ labels the
additional dimensions. 
(We have dropped Lorentz indices on the 
fluctuations  for simplicity.) General fluctuations can  be written 
as superpositions of modes, $\hat{h} = e^{i p.x} \hat{\psi}(z)$, where
$\hat{\psi}$ is an eigenmode of the following equation
in the extra dimensional coordinates: 
\begin{equation}
\label{eigenvalue}
\left(-\partial_j \partial^j + V(z)\right)\hat{\psi}(z)=-m^2 \hat{\psi}(z),
\end{equation}
and $p^2 = m^2$. 
This implements the Kaluza-Klein (KK) reduction of
 the higher-dimensional gravitational fluctuations in terms of 
four-dimensional KK states, with mass-squared, $m^2$, 
given by  the eigenvalues of Eq. (\ref{eigenvalue}).
%, and with fixed 
%wave-function in the extra dimension, $\hat{\psi}(z)$. 
It is useful to note that
Eq. (\ref{eigenvalue}) takes the form of an analog non-relativistic quantum 
mechanics problem.
If there is a zero-mode (which  is guaranteed if
the background preserves four-dimensional Poincare-invariance) which is also a
  normalizable state in the spectrum of Eq. (\ref{eigenvalue}), it
is the wave function associated with the  four-dimensional graviton.
This state is indeed a bound state and falls off rapidly
away from the brane.

In addition there exists  a tower of  KK
modes. If there were a gap, as is conventional in  product
space compactifications, one would reproduce four-dimensional 
gravity up to the scale determined by the gap.
Instead, in our theory, there is a continuous
KK spectrum with no gap. However, four-dimensional
physics is extremely well approximated 
because the bound state mode reproduces 
conventional four-dimensional
gravity, while   the other KK modes give only a small correction,
as we will demonstrate.
 
The set-up for our  theory is a single 3-brane
with  positive tension,
embedded in a five-dimensional bulk spacetime.  
In order to carefully quantize the system, and treat the
non-normalizable modes which will appear in the Kaluza-Klein
reduction, we choose to first work in a finite volume by
introducing another brane at a distance $\pi r_c$ from the
brane of interest, and taking the branes to be the boundaries of a finite
fifth dimension. We will eventually take this second 
brane to infinity,
thereby removing it from the physical set-up. Analogous domain walls were 
discussed in  Ref. \cite{cvetic} and references therein.
%Our regulated set-up is therefore
%the same as in ref. \cite{} but  exchanging the roles of the visible sector
% and hidden sector branes. 
The action for our system is
\begin{eqnarray}
\label{action}
S &=& S_{gravity} + S_{brane} + S_{brane'} \nonumber \\
S_{gravity} &=& 
\int d^4 x \int d y \sqrt{-G} \{- \Lambda + 2 M^3 R \} 
\nonumber \\
S_{brane} &=& \int d^4 x \sqrt{-g_{brane}} \{ V_{brane} + {\cal L}_{brane} \},
\end{eqnarray}
where $R$ is the five-dimensional 
Ricci scalar made out of the five-dimensional metric, $G_{MN}$,
and $\Lambda$ and $V_{brane}$ are cosmological terms
in the bulk and boundary respectively. 
The coupling to the branes and their fields and the related orbifold boundary 
conditions are described in Refs. 
\cite{us}(see also \cite{me1}). 
(The new coordinate $y$ is $r_c \phi$ in the coordinates of Ref. 
\cite{us}.)
%The bulk metric is $G_{MN}(x)$, the four dimensional
%induced metric on the brane is $g_{\mu \nu}(x)=G_{MN}(Y(x))\partial_\mu Y^M 
%\partial\nu Y^N$, where $Y^\mu(x)$ parameterizes the location
%of the brane. The associated action is
%\begin{eqnarray}
%\label{action}
%S &=& S_{gravity} + S_{vis} + S_{hidden} \nonumber \\
%S_{gravity} &=& 
%\int d^4 x \int_{- \pi}^{\pi} d \phi \sqrt{-G} \{- \Lambda + 2 M^3 R \} 
%\nonumber \\
%S_{brane} &=& \int d^4 x \sqrt{-g_{brane}} \{ V_{brane} + {\cal L}_{brane} \}.
%\end{eqnarray}
%In order to carefully quantize the system, and treat the
%nonnormalizable modes which will appear in the Kaluza-Klein
%reduction, we choose to first work in a finite volume by
%introducting another brane at a distance $r_c$ from the
%brane of interest. We will eventually take this brane to infinity,
%thereby removing it from the physical set-up.
%The action for this brane is
%\begin{equation}
%S_{brane'} = \int d^4 x \sqrt{-g_{brane'}} \{ V_{brane'} + {\cal L}_{brane'}
%\end{equation}

The solution to Einstein's equations was
derived in Ref. \cite{us} and is
\begin{equation}
\label{solution1}
ds^2 = e^{- 2 k |y|} \eta_{\mu \nu} dx^{\mu} dx^{\nu} + d y^2, 
\end{equation}
where $0 \leq y \leq \pi r_c$ is the extra-dimensional coordinate and $r_c$ 
is essentially a compactification ``radius''. 
It  can
be identified as a slice of  $AdS_5$. The solution holds only when the 
boundary and bulk cosmological terms are related by
\begin{equation}
\label{conditions}
V_{brane} = - V_{brane'} = 24 M^3 k, ~~ \Lambda = - 24 M^3 k^2,
\end{equation} 
which we assume from now on.  We remind the reader
that this condition amounts  to setting the cosmological constant
of the four-dimensional world to zero in this context, and we simply
accept this necessary fine tuning without
further explanation here.
 Notice that in the solution given in Eq. (\ref{solution1}),
we have reversed  the labels for   the ``visible'' and ``hidden'' branes 
relative 
to Ref. \cite{us}. The solution for the background
metric is the same, with the metric exponentially
falling from one brane to the other. However, 
whereas in the solution to the hierarchy problem proposed in Ref \cite{us}
the massless graviton wavefunction is biggest on
the hidden brane, in the scenario
considered here it is critical that
the graviton is ``bound'' to
the visible brane. 

We also review the derivation of the four-dimensional effective 
Planck scale, $M_{Pl}$  as given in Ref. \cite{us}. The four-dimensional 
graviton 
 zero mode follows from the 
solution, Eq. (\ref{solution1}), by replacing the Minkowski metric by a  
four-dimensional metric,  $\overline{g}_{\mu \nu}(x)$. It is described 
by an 
effective action following from substitution into Eq. (\ref{action}), 
\begin{equation}
\label{effaction}
 S_{eff} \supset \int d^4 x \int^{\pi r_c}_{0} d y ~
2 M^3 r_c e^{-2 k  |y|} \sqrt{\overline{g}}~ \overline{R},
\end{equation}
where $\overline{R}$ denotes the four-dimensional Ricci scalar
 made out of 
$\overline{g}_{\mu \nu}(x)$, in contrast to the five-dimensional Ricci 
scalar, $R$, made out of $G_{MN}(x, y)$.
Because the effective field is four-dimensional, we can 
explicitly perform the $y$ integral to obtain a purely four-dimensional 
action. From this we derive
\begin{equation}
\label{effplanck}
M_{Pl}^2 = 
2 M^3  \int_{0}^{\pi r_c} dy  e^{-2 k |y|} = \frac{M^3}{k}
[1 - e^{- 2 k r_c \pi}].
\end{equation}
We see that there is a well-defined
value for $M_{Pl}$, even in the $r_c \to \infty$ limit.
This is a clue that one can get a sensible effective
four-dimensional theory, with the usual Newtonian
force law, even in the infinite radius limit, in sharp
contrast to the product-space expectation that $M_{Pl}^2 = M^3 r_c \pi$.

%In fact, in this limit, we do not need to impose
%orbifold boundary conditions; we can eliminate
%the second brane altogether. In this case,
%we will write the solution to Einstein's equations as
%\begin{equation}
%ds^2=e^{-2k |y|} \eta_{\mu \nu} dx^\mu dx^\nu +dy^2
%\end{equation}
%where $k$ is defined as in Eq. \ref{keqn}.

Clearly, there  is no problem with taking the $r_c\to \infty$
limit of the background metric given above. 
This will remove the ``regulator'' brane 
from the set-up. However, we still
need to determine whether the spectrum of {\it general} linearized 
tensor 
fluctuations $G_{\mu \nu} = e^{-2 k |y|} \eta_{\mu \nu} + h_{\mu \nu}(x,y)$ 
is consistent with four-dimensional experimental gravity.
This requires an understanding of all modes that
appear in the assumed four-dimensional effective theory.
We therefore  perform a Kaluza-Klein reduction
down to  four-dimensions. 
%In order
%to carefully quantize the modes, we return to the
%finite volume limit, after which we will
%again take the $r_c \to \infty$ limit. 
%For finite $r_c$, we need to introduce boundary
%conditions. We take these to be the same as in Ref. \cite{us},
%namely the orbifold $S^1/Z_2$, implying periodicity
%in our solution as a function of $y$ and that we
%retain only symmetric wave functions of $y$.
To do this, we need to do a separation
of variables; we write
$h(x,y)=\psi(y)e^{i p \cdot x}$, where $p^2=m^2$
and $m^2$ permits a solution to  the linearized equation of motion for tensor 
fluctuations following from Eq. (\ref{action}) expanded about Eq. 
(\ref{solution1}): 
\begin{equation}
\label{linearized}
\left[{-m^2 \over 2} e^{2 k |y|}  -{1 \over 2} \partial_y^2 
-2k \delta(y) +2k^2 \right ]\psi(y)=0,
\end{equation}
where the assumed orbifold 
 boundary conditions tell us to  consider only even functions of $y$.
 The effect of the 
regulator brane will be considered later;  here it has been taken to infinity.
The $\mu \nu$ indices are the 
same in all terms if we work in the gauge where $\partial^{\mu} 
h_{\mu \nu} = h^{\mu}_{\mu} = 0$, so they are omitted.
Here $m$ is the four-dimensional mass of the KK excitation. 

It is more convenient to put the above equation into the form of an analog 
 non-relativistic quantum mechanics problem by making a change of
variables, $z \equiv sgn(y) \left( e^{k |y|} -1 \right)/k$,  $\hat{\psi}(z) 
 \equiv \psi(y) e^{k |y|/2}$, $\hat{h}(x,z) \equiv h(x,y) e^{k |y|/2}$.  
Eq. (\ref{linearized}) then reads
%\begin{equation}
%{\cal S }=\int d^4 x
%\int {dz \over k} \hat{h} \left[-{\partial_\mu^2 \over 2}+k^2 H(z) 
%\right ] \hat{h},
%\end{equation}
%where 
\begin{equation}
\left [-{1 \over 2}\partial_z^2 +V(z) \right ] \hat{\psi}(z) = m^2 \hat{\psi},
\end{equation}
where 
\begin{equation}
V(z)={15 k^2 \over 8 (k|z|+1)^2} -{3 k \over 2} \delta (z).
\end{equation}
Much can be understood from the general shape of this analog non-relativistic 
potential.

First, the $\delta$-function supports a single
normalizable bound state mode;  the
remaining eigenstates are continuum modes. We have
already discussed the role of the bound state
mode as the massless graviton of the effective four-dimensional
theory. With the explicit form of the KK ``potential'',
we can also understand the properties of the continuum modes.
First, since
 the potential falls off to zero
as $|z| \to \infty$, there is no gap, and the continuum modes asymptote to 
plane waves. Furthermore, the amplitudes of the continuum modes
is suppressed near the origin, due to the potential barrier
near $z=0$. Finally, 
the  continuum  KK states have all possible $m^2 > 0$. 

The precise continuum modes are given in terms of Bessel functions,
and are a linear combination of $(|z| +1/k)^{1/2} Y_2(m  (|z|+1/k))$
and $(|z| +1/k)^{1/2} J_2(m  (|z|+1/k))$.  The zero mode wavefunction 
follows (after 
changing variables) from Eq. (\ref{solution1}), 
$\hat{\psi}_0(z) = k^{-1} (k |z| + 1)^{-3/2}$.\footnote{
Though the zero mode is not a Bessel function, it
 is the limit  of $m^2   (|z| +1)^{1/2} Y_2(m  (|z|+1))$ 
when $m \to 0$.}
We can better understand the KK modes
by studying the small and large argument limits of the
Bessel functions. For small ${m \over k}(|z| +1)$ we 
have 
\begin{equation}
\label{small}
J_2(m (|z| +1/k)) \sim {m^2 (|z| + 1/k)^2 \over 8}, \ \ 
Y_2(m (|z| +1/k)) \sim - {4 \over  \pi 
m^2 (|z| + 1/k)^2} - {1 \over \pi} 
\end{equation} 
Therefore to satisfy the boundary condition implied by the $\delta$-function 
potential on the brane at $z = 0$, for small $m$ (relevant at long 
distances) we must choose the 
linear combination,
\begin{equation}
\label{eigen}
\hat{\psi}_m \sim N_m
(|z|+1/k)^{1/2}\left  [Y_2(m  (|z|+1/k))  +  {4 k^2 \over \pi m^2}
J_2(m  (|z|+1/k)) \right ]. 
\end{equation}
Here $N_m$ is a normalization constant.
For large $mz$,
\begin{equation}
\label{largez}
\sqrt{z} J_2(m z) \sim \sqrt{{2 \over \pi m }} \cos(mz -{5 \over 4} \pi), 
~~~~ 
\sqrt{z} Y_2(m z) \sim \sqrt{{2 \over \pi m }} \sin(mz -{5 \over 4} \pi).
\end{equation}
Let us now consider what happens when we reintroduce the regulator brane at 
$y_c \equiv \pi r_c$, that is $z_c \equiv (e^{k  \pi r_c} - 1)/k$. 
It simply corresponds to  a new  boundary condition at $z_c$, 
\begin{equation}
\label{bc}
\partial_z \hat{\psi}(z_c) = - {3 k \over 2 (k z_c +1)} \hat{\psi}(z_c).
\end{equation}
It is easy to check that our zero-mode satisfies this new boundary  condition. 
However, this condition does restrict the allowed continuum
modes and quantizes the allowed values of $m$. 
For large $z_c$ they are all in the plane-wave 
asymptotic regime of Eq. (\ref{largez}) 
when they satisfy the new condition. Therefore their masses are approximately 
 quantized in units of $1/z_c$. Furthermore their normalization constants are 
predominantly those of plane waves, in particular, $N_m \sim \pi m^{5/2}/(4 
k^2 \sqrt{z_c})$.

Having obtained the  large but finite $r_c$ asymptotics we can determine the 
proper measure for sums over the continuum states in the $r_c \to \infty$ 
limit. Because these asymptotics were dominated by plane wave behavior, this 
measure is simply  $dm$ after dropping the $1/\sqrt{z_c}$ factor in $N_m$ 
to go a continuum normalization. We have also demonstrated the claim made in 
 Ref. \cite{us}, that when $z_c$ 
is kept large but finite, 
these KK states are quantized in units of $1/z_c$, 
which in Ref. \cite{us}  corresponded to the TeV scale. Also note that 
the normalized KK wavefunctions at the brane at $z_c$ are all of 
order $1/\sqrt{z_c}$ since they are all plane waves at a maximum or minimum
 according 
to Eq. (\ref{bc}), which is $k z_c$ times larger than $\hat{\psi}_0(z_c)$. This
proves the claim of Ref. \cite{us} that the KK states couple 
$10^{15}$ more strongly to matter on the brane at $z_c$ than does the massless
 graviton. 

Clearly, the $r_c \to \infty$ limit
gives rise to a theory with   a semi-infinite extra dimension. 
However, we were initially interested in studying a theory
without the orbifold boundary condition, that is 
a theory with infinite extent in both the positive and negative $z$
direction. It
is trivial to extend the set-up
we have studied  to a fully infinite extra dimension by  
allowing even and odd functions of $z$ rather than the restriction to purely 
even functions demanded by the orbifold conditions. From now on we will 
consider this to be the case. However, we will make use of
the density of states we have found by the study of the finite
volume situation.

%modes will be quantized in units
%of $1/z_c$, where $z_c=e^{ k r_c}$. Restoring the mass scale
%$m$ turns this into $k e^{- k r_c}$. As $r_c \to \infty$,
%one finds modes with $m \to 0$ which is the continuum
%we discussed earlier.  This result also confirms
%the result of our previous paper, namely that
%the modes in that model were quantized in units of TeV,
%which in that paper was $k e^{-k r_c}$.   

%We can also study the small $mz$ limit.  Here, we 
%have 
%\begin{equation}
%J_2(m z) \to {m^2 z^2 \over 4}, \ \ Y_2(mz) \to 1{1 \over 4 \pi z^2} 
%\end{equation} 
%For small $m$, the normalized solution that satisfies
%the boundary condition at $0$ takes the form
%\begin{equation}
%\hat{h}\sim {m^3 \over k^3}
%\sqrt{|z|+1} \left{ Y_2({m \over k} (|z|+1))+{4 k^2 \over \pi m^2}
%J_2({m \over k} (|z|+1)),
%\end{equation}
%where the prefactor is an approximate normalization factor.

Having found the KK spectrum of the effective four-dimensional
theory, we can  now  compute the non-relativistic gravitational 
potential between two particles of mass
$m_1$ and $m_2$  on our brane at $z = 0$, 
that is the static potential generated by
exchange of the zero-mode  and continuum Kaluza-Klein  mode propagators.
It is
\begin{equation}
V(r) \sim G_N { m_1 m_2 \over r} +\int_0^{\infty} dm { G_N \over k} 
{ m_1 m_2 e^{- m r} \over r} {m \over k}.
\end{equation}
Note there is a Yukawa exponential suppression in the massive Green's functions
for $m > 1/r$, and the extra power of $m/k$ arises from the suppression of 
continuum wavefunctions at $z = 0$ following from Eq. (\ref{small}),
 discussed
 above. The coupling $G_N/k$ in the second term is nothing but the fundamental
 coupling of gravity, $1/M^3$, by Eq. (\ref{effplanck}).
 Therefore, the potential behaves as
\begin{equation}
V(r) = G_N { m_1 m_2 \over r}\left(1+{1 \over r^2 k^2} \right)
\end{equation}
This is why our theory produces an effective four-dimensional
theory of gravity . The leading term due
to the bound state mode is  the usual Newtonian potential;
the KK modes generate an extremely
suppressed correction term, for $k$ taking the
expected value of order the
fundamental Planck scale and $r$ of the size
tested with gravity.
Furthermore, since our propagators are relativistic in general,  
going beyond the non-relativistic approximation we find all the proper 
relativistic corrections, again with negligible corrections from the continuum
modes. 

 From the small $m$ limit of the continuum wave
functions, we also  
learn that the production of the 
the continuum gravitational  modes from the brane at $z=0$ is suppressed by 
$(dm/k) 
(m/k)$ due to the continuum
wavefunction suppression there.
This is very important, because it means the amplitude
to produce the continuum  modes in low-energy processes on the brane 
is extremely small, far smaller than gravitational strength.
Were this not the case, we would be in danger of continuously losing
energy to the additional dimension. Because of this suppression
factor, the probability of producing KK modes is
suppressed by $(p/k)^2$ relative to the zero mode, 
where $p$ is the momentum of a process.
For $k$ of order the Planck scale, this is extraordinarily small
for any process we presently observe, or are ever likely to observe.

We have shown that a scenario
with an infinite fifth dimension 
in the presence of a brane can generate
a theory of gravity which mimics 
purely four-dimensional gravity, both with
respect to the classical gravitational
potential and with respect to gravitational radiation.
It is also important to verify that the
gravitational self-couplings are not
significantly modified.
For gravitions which couple to matter
fields with {\it gravitational} strength,
these 
have been tested at the
 $10^{-3}$ level of precision (see Ref. \cite{damour}
for a review).
 Because the KK modes have $p/k$-{\it suppressed}
coupling to matter on the brane relative to the zero mode,
tests in which the KK modes ultimately couple
to matter on the brane (ie detectors) are insensitive
to the existence of the many new gravitational
modes. It is only the zero mode which contributes
at any measurable level.
 The zero mode exchanges and self-couplings are just those of a 
four-dimensional general relativistic dynamics
 described by Eq. (\ref{effaction}).

%Finally, having explicitly worked out the Kaluza-Klein
%modes, we can verify the claim made in \cite{rs2}
%that in the theory with two branes where the visible
%world is the negative tension brane at $\phi=\pi$
%that the Kaluza-Klein modes couple
%with strength ${1 \over TeV}$.  
%Here, there is a coupling
%$\sqrt{-g} g^{\mu \nu} T_{\mu \nu}$. Extracting the
%factor $e^{-4 kr_c}$ from the metric determinant,
%$e^{ 2 k r_c}$ from the explicit metric factor, $e^{2kr_c}$
%from expanding the metric to get the coupling of the KK mode $h$,
%and normalizing the fields appearing in $T_{\mu\nu}$ so that
%the kinetic energy terms are canonical introducing a factor of $e^{2k r_c}$
%leaves an overall factor of $e^{2k r_c}$. When we expand $h$
%in terms of the KK modes, the wave function multiplying the
%zero mode scales as $e^{-2kr_c}$, whereas the wave function
%for the nonnormalizable modes at the second boundary scales as $e^{-k r_c}$.
%Therefore, the coupling of the true graviton is suppressed by
%$M_{Pl}$, whereas the couplings of the remaining continuum
%modes are suppressed by $TeV$.

However, it is important to also verify that
the energy loss induced by gravitational
self-interactions is also insignificant,
that is the coupling of the zero mode 
to KK modes which do {\it not} ulimately
couple back to matter on the brane. We
will now demonstrate that this
is also small. However, to do so
requires an understanding of the limitations
of the perturbative approach to the
graviton fluctuations that  we have implicitly assumed.

We will now show that for any finite energy,
the graviton self-coupling gets
large at an energy-dependent value of
the coordinate $z$. Subsequently,
we will demonstrate that fluctuations 
originating on the brane in low-energy processes have
only a small probability to
get to this large $z$. 
Graviton emission and the
associated missing energy can
be bounded within the framework
of the low-energy effective theory,
and we will argue that it is small.

%However, there is still a potential worry.  The
%origin of this apparent problem  is the fact
%that we have a gapless continuum, so that
%in any energy interval, no matter how low,
%5there is an infinite number of modes. By
%general covariance, each of these modes
%couples to the massless graviton with gravitational
%strength.  So in the decay of an off-shell
%5graviton, or in loop corrections to  the gravitational
%potential, there appears to be a divergence
%arising from summing over modes. Explicitly,
%\begin{equation}
%S \supset \int d^4 x {\hat{h}_0^{\mu \nu}(x) \over M_{Pl} }
%\int {dm \over k} \partial_\mu \hat{h}_m^{\alpha \beta}(x) \partial_\nu
%\hat{h}^m_{\alpha \beta}(x)
%\end{equation}
%However, 
%this is a red herring, as we now demonstrate.

We first explore the limitation of the
linearized KK spectrum calculation we have performed. 
The problem can best be understood by
considering the graviton couplings in position space in the extra dimension.
We have solved for the KK
modes in a linear expansion about the classical
gravitional background. The linear approximation
is well justified in the vicinity of the
brane, but higher order terms in the perturbation
are significant  far from the brane.

The leading term giving the graviton coupling
is
\begin{equation}
\label{strong}
S \supset \int d^4 x \int dz  \sqrt{-G} R \supset
\int d^4 x \int dz z^{3/2}~
{\hat{h}^{\mu \nu}(x,z) \partial_\mu \hat{h}_{\alpha \beta}(x,z)
 \partial_\nu \hat{h}^{\alpha \beta}(x,z) \over M_{Pl}}.
\end{equation}
In this language, it is clear that there
is a strong coupling at large $z$.   The source of the problem
is that the coupling
of the gravitons diverges as one approaches the AdS horizon.
This is related to the standard blue-shift
near the horizon of AdS space. This can be seen by examining the large-$z$ 
behavior of our continuum KK modes from Eq. (\ref{largez}), 
$\hat{\psi}(z) \sim e^{i mz/k}$. No matter how soft such a mode is when
 produced near the 3-brane (small $m$), it is blue-shifted into a hard mode 
 as measured in  the background five-dimensional metric at large $z$, $
G_{MN} \sim \eta_{MN}/z^2$, $M, N = \mu, z$ . 
Therefore at large $z$ it can have strong 
gravitational couplings to other modes.

In fact, this strong coupling in our 
effective field theory description is probably essential if one is to obtain
a consistent four-dimensional theory. This is because our theory
respects four-dimensional general covariance and therefore
each graviton mode in the four-dimensional effective
theory must be coupled consistently. However, we know
the wave function overlap between the zero mode and
the continuum modes is very small.  In position
space in the five-dimensional theory, this must be
compensated by a large coupling of the linearized modes at large $z$ 
to reproduce the
correct four-dimensional result.

The question remains whether this strong-coupling regime is problematic. 
We presume that if our set-up is embedded in  fundamental string theory, then 
the strong coupling interactions at large $z$
has an alternative description not encompassed
by our perturbative approach. Essentially all the physics relevant
to the four-dimensional world occurs on or near the
brane. 
 If strong  interactions were a very frequent conseqeunce of 
low-energy processes on the 3-brane however, we would lose all predictivity 
within the effective field theory framework of this letter. Fortunately this is
 not the case, as we show by a simple estimate.  
By Eq. (\ref{strong}), five-dimensional gravitational fluctuations have 
couplings of order  
Energy$.z^{3/2}/M_{Pl}$.  We presently only perform experiments at energies 
of order a TeV or less.
In order for gravitational fluctuations emerging from such processes to be 
coupled more strongly than the percent level for example,  they must escape 
out to 
$z > 10^{9}$. Now, as we saw, there is negligible probability for low-energy
brane processes to directly create continuum KK modes, they will almost 
exclusively result in massless gravitons (possibly off-shell). So our problem 
reduces to finding the probability for these massless gravitons to be at 
 $z > 10^{9}$. Consulting the massless graviton's wavefunction, this 
probability is $10^{-18}$! Thus, while low-energy brane processes {\it can} 
lead to strong gravitational interactions at large $z$, not captured by 
weakly coupled effective field theory, they almost never do. With almost unit 
probability, low-energy brane processes interact with the massless graviton 
mode according to an effective four-dimensional general relativity, without 
exciting continuum KK modes.

%The
%effective field theory approach would argue that
%it is, because essentially all the physics relevant
%to the four-dimensional world occurs on or near the
%brane.  This can be made explicit by
%calculating the rate of energy transfer across
%some artificial boundary placed sufficiently
%far from the brane. If one can find a location
%$z_0$ where both the energy transfer across it
%and the graviton couplings are sufficienlty
%small, one can restrict attention to the
%effective field theory defined in the region
%$z < z_0$. Explicitly, one finds
%the probability that the continuum graviton
%is found at a location $z>z_0$ is
%$P(z>z_0)\sim  \left({1 \over z_0}\right)^2$
%whereas the coupling of the zero mode to the
%graviton modes of momentum $p$  behaves as $p/k z_0^{3/2}$.
%Both of these are much less than one if
%$1<<z_0<<10^{10}$, which is clearly not difficult to achieve.

%What justifies the restriction to the region $z<z_0$?
%In essence, we know the description of the theory
%must change near the horizon, assuming
%we are describing a consistent low-energy
%effective field theory.  This could be due
%to the fact that the couplings soften,
%or because the description changes. One
%does not expect the effective field theory
%to apply beyond this regime. Fortunately,
%physical results at low-energy depend
%extremely weakly on physics in this region,
%restoring the utility of the effective
%field theory description.

To conclude, we have found that we can
consistently exist with an infinite fifth
dimension, without violating known tests
of gravity. The scenario consists of
a single 3-brane, (a piece of) $AdS_5$ in the bulk,
and an appropriately tuned tension
on the brane. The need for this delicate adjustment 
is the equivalent of the 
the cosmological constant problem 
in this context, and is taken as a given
and not solved.

In this framework, we have found that
an inevitable consequence is a bound
state graviton mode, whose shape is
determined by the brane tension and bulk cosmological constant. 
There are no very large or small numbers
assumed for the different gravitational mass scales
in the problem, so the four dimensional
Planck scale is comparable to the 
fundamental mass scale of the higher
dimensional theory. In addition
to the bound state mode, there is
a continuum of Kaluza-Klein modes. These
have very weak coupling to low-energy states on
the brane, but are essential to
the consistency of the full theory
of gravity and would couple strongly to Planck-energy brane processes.

Notice that one interpretation of
our result is as a solution
to the moduli problem, for the
particular modulus determining the
distance between two branes. It
says that the usual disasterous possibility, namely
 that the modulus runs away to infinity,
is perfectly acceptable. Furthermore,
in the $r_c \to \infty$ limit,
the modulus is not coupled to
matter on the brane, and the
need for a modulus mass is eliminated.
In fact, it is interesting to speculate
that the problem associated with
geometric moduli can be eliminated,
because there is no compactification at all,
and gravity is bound by a mechanism
akin to that suggested in this letter.
 This
would be a worthwhile problem to
study within the context of fundamental
string theory.

Because our effective theory clearly breaks
down before $r_c \to \infty$ for (rare) 
processes involving Kaluza-Klein modes, it
might be thought that this theory
is in some sense compactified. It
should be noted that this theory is
clearly very different from truly
compactified theories. First, the extra-dimensional 
measure is straighforwardly $dy$ 
in the coordinate $y$, which takes values on the entire real line. The
low dimensional Planck scale and
all physical parameters of the effective
four-dimensional theory are independent
of $r_c$, so long as it is
much greater than $1/k$. Furthermore,
at sufficiently low energies, the
theory breaks down further and further
from the origin, since the blue
shifted energy becomes large further
away.  From these perspectives,
the theory provides a well defined
alternative to geometric compactification.
However, it is intriguing to speculate
that there exists  a dual description
of this theory in terms of a cut-off conformal
field theory in four-dimensions, akin
to the duals discussed in \cite{maldacena}.

Many interesting questions remain to be addressed.
Given a valid alternative to conventional
four-dimensional gravity, it is important
to also  consider the astrophysical and cosmological
implications. This different scenario
might even  provide a new perspective
for solving uresolved issues
in quantum gravity and cosmology.

~

{\bf Acknowledgements:} We wish to acknowledge
useful discussions
with Vijay Balasubramanian, Martin Gremm,  Gary
Horowitz, Igor Klebanov, Juan  Maldacena,  Joe Polchinski,
Massimo Poratti,
 Herman Verlinde,  Dan Waldram, and especially Andy Strominger. 
The research of Lisa Randall
was supported in part by DOE under cooperative
agreement DE-FC02-94ER40818 and
under grant number DE-FG02-91ER4071. The research of Raman Sundrum was 
supported by
the US Department of Energy under grant no. DE-FG02-94ER40818.

\end{document}